\documentclass[twocolumn,aps,prb,floats]{revtex4-2}

\usepackage{amsmath,amssymb,bm,siunitx,graphicx}
\DeclareMathOperator{\tr}{Tr}

\newcommand{\femv}{\bm f^\mathrm{em}}
\newcommand{\felv}{\bm f^\mathrm{el}}
\newcommand{\upd}{\mathrm{d}}

\begin{document}

\begin{abstract}
We investigate the space-time evolution of the magnetization induced in a 2D ferromagnetic island by a short pulse of a rotating electric field. The field generates elastic twists that act on the magnetization via the Barnett effect: magnetization by rotation.  Analytical studies are conducted within classical electrodynamics of continuous media and continuous elastic theory, while numerical studies are performed using discretized Landau-Lifshitz spin dynamics on the atomic lattice. The effect is studied for typical parameters of magnetic oxides at various field frequencies and amplitudes, and for various strengths of magnetic anisotropy, exchange, and damping. The possibility of reversing the island magnetization with an electric-field pulse is demonstrated.
\end{abstract}

\title{Barnett effect generated by a rotating electric field in a ferromagnetic film}
\author{ Jorge F. Soriano and Eugene M. Chudnovsky}
\affiliation{Physics Department, Herbert H. Lehman College and Graduate School, The City University of New York, 250 Bedford Park Boulevard West, Bronx, New York 10468-1589, USA }
\date{\today}
\maketitle

\section{Introduction}

Magnetization reversal by the electric field or current in memory devices has been a paradigm of modern magnetism. It has been pursued because of slow inductance effects associated with switching the magnetic moment by the magnetic field. The electric field or current can be generated much faster. However, a linear relation between the magnetization and the electric field or the charge current, ${\bf M} \propto {\bf E}$ or ${\bf M} \propto {\bf J}$, is prohibited by symmetry. In microelectronics, the focus has been on multiferroic materials, in which magnetization is coupled to spontaneous electric polarization, and on switching the magnetic moment with a spin-polarized electric current \cite{Fert-RMP2024}.  Voltage-controlled magnetic anisotropy has been another common approach for developing nonvolatile magnetic memory \cite{Nozaki-2019}.

In recent years, there has also been noticeable research on switching spins by mechanical rotations, using the fact first observed by Barnett \cite{Barnett} that a linear relation between the magnetic moment and angular velocity, ${\bf M} \propto  {\bm \Omega}$, is permitted by symmetry. In the rotating coordinate frame rigidly coupled to the crystal lattice, the spins feel the effective magnetic field, which is the essence of the Larmor theorem \cite{Larmor}. The Barnett effect at the nanoscale has recently been demonstrated by Davies et al. \cite{Davies-Nature2024} through flipping spins in a paramagnetic substrate by circularly polarized phonons. 

In the past, the Barnett effect and the reciprocal Einstein--de Haas effect \cite{Einstein1,Einstein2} were applied to interpreting experiments with nanocantilevers  \cite{Kovalev-APL2003,Kovalev-PRL2005,Wallis-APL2006,JCG-PRB2009,Mori-PRB2020}. They were invoked for understanding the mechanism of spin relaxation and decoherence due to interaction of spins with chiral phonons \cite{EC-PRB2002,EC-PRL2004,EC-DG-PRL2004,EC-DG-RS-PRB2005,Dornes-Nature2019,Tauchet-Nature2022}, and for explaining experiments  with individual magnetic molecules that are free to rotate \cite{EC-PRL1994,EC-DG-PRB2010,DG-EC-PRB2021,Wernsdorfer-2015}. Most recently, it was demonstrated that fast-moving dislocations generated by a strong elastic stress are capable of flipping individual spins \cite{EC-JS-PRB2025} as well as the magnetization in a ferromagnetic film \cite{EC-JS-EPL2026} via the Barnett effect. 

The Barnett and Einstein--de Haas effects may explain large effective magnetic fields from high-frequency chiral phonons reported in $4f$ paramagnets \cite{Juraschek-PRR2022} and rare-earth halides \cite{Luo-Science2023}. They may provide a complementary angle at numerous experiments that reported flipping of spins in solids by applying short electric pulses \cite{Yang-Science2017}, laser  \cite{Xu-JMMM2022}, microwave  pulses \cite{Cai-PRB2013,Miyashita-PRL2023}, and surface acoustic waves \cite{Tejada-EPL2017,Camara-PRA2019}. At the microscopic level, these phenomena correspond to the transfer of the phonon angular momentum  \cite{Zhang-PRL2014,DG-EC-PRB2015,Nakane-PRB2018,DG-EC-PRB2021} to the atomic spin. 

A nanoelectromechanical system (NEMS) based upon the Barnett effect was proposed in Ref.\ \onlinecite{EC-RJ-JAP2015}. It considered a small multiferroic particle attached to a torsional cantilever or a monodomain ferromagnetic particle attached to a ferroelectric cantilever. A short electric field pulse applied to such a NEMS would generate its fast rotation, which would produce the effective magnetic field in the coordinate frame of the particle, leading to the magnetization reversal. 

\begin{figure}\centering
\includegraphics[width=0.95\linewidth]{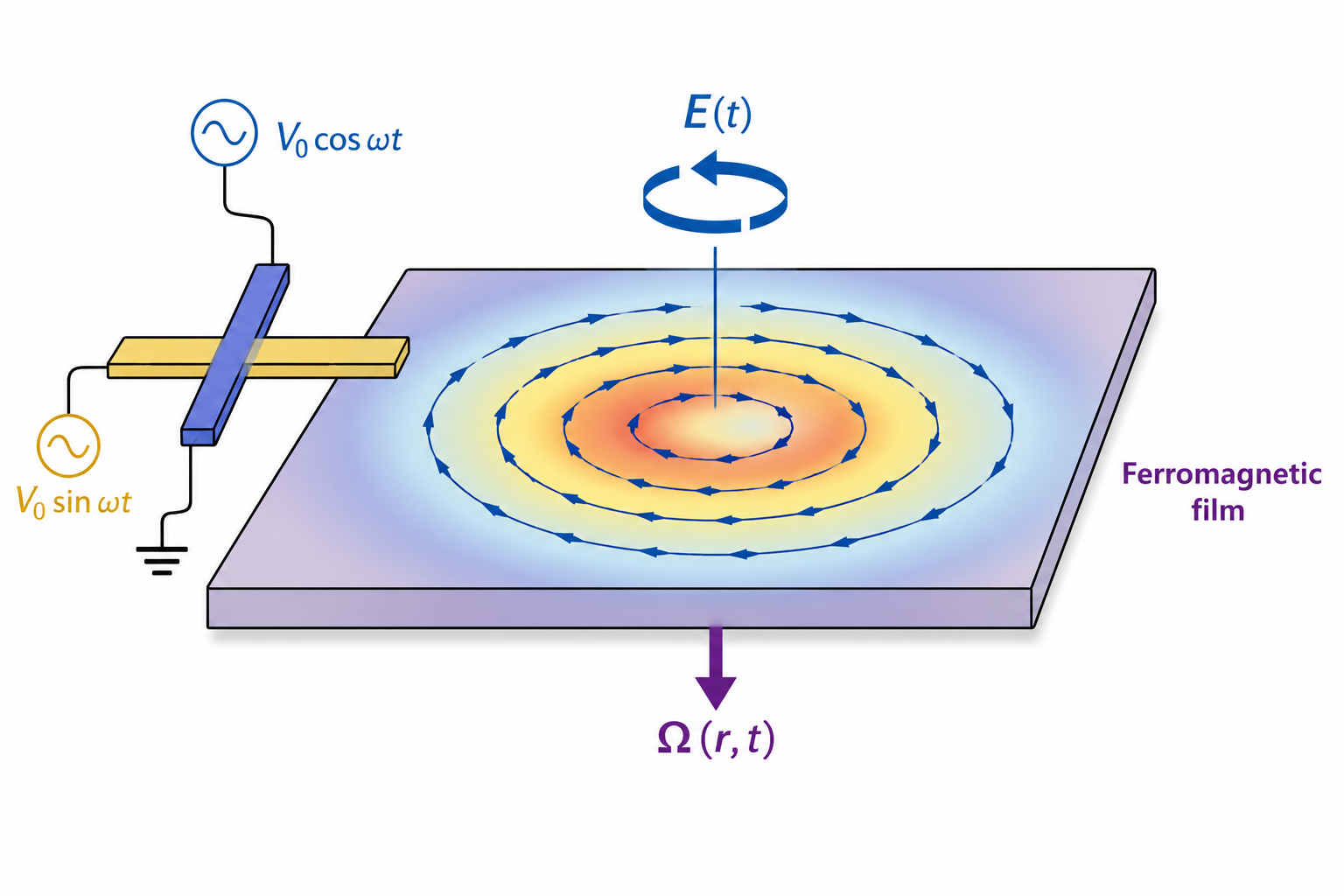}
\vspace{-1cm}
\caption{Barnett effect produced in a ferromagnetic film by a rotating electric field.}
\label{setup}
\end{figure}
In this paper, we investigate the possibility of electric-field-induced magnetization reversal in a 2D island of ferromagnetic material without invoking a mechanical rotor or a cantilever. Instead of a rigid body rotation, we consider a distributed Barnett effect in a crystal lattice induced by a short pulse of a rotating electric field produced near the film by nano-scale electrodes, see Fig.\ \ref{setup}. The coupling with the lattice is mediated by the Maxwell stress \cite{LL-ECM}. To avoid rapid screening, we chose typical parameters of a weakly conducting magnetic oxide and compute the elastic deformation produced in the film by the electric field of a chosen profile. This allows us to obtain the space-time profile of the elastic twist and the corresponding effective magnetic field acting on the spins. We then solve the system of coupled Landau--Lifshitz equations for the spins and investigate the range of parameters that permit the magnetization reversal.

The paper is organized as follows. The model is formulated in Section II. It outlines the steps used to calculate the space-time dependence of the elastic twist induced by a rotating electric field, and the spin dynamics that results from it via the Barnett effect. The deformation fields are computed in Section III, which also provides the condition under which elastic twists can produce a significant impact on the magnetization. The dependence of the effect on the frequency of the rotating electric field is discussed in Section IV. Numerical results on the magnetization dynamics are presented in Section V, which considers two regimes corresponding to low and high damping. An analytical solution of the Navier--Cauchy equation for the deformations generated by the rotating field is given in Appendix A. The final Section VI contains a brief discussion of the results.

\section{The model}
We intend to illustrate the magnetic effects that a rotating electric field has on a magnetic lattice via the local, time-dependent deformations it produces on the lattice, which induce non-inertial effects on the evolution of the spin at each lattice site. Our description of the model is thus separated into four aspects: the electric field, its coupling to the lattice, and the elastic and magnetic behaviors of the lattice.

We consider an electric field rotating in the $x-y$ plane with angular frequency $\omega_0$ (frequency $f_0=\omega_0/2\pi$), which we write as
\begin{equation}
	\bm E=w(t)f(r)E_0\left<\cos(\omega_0t),\sin(\omega_0t),\varepsilon_z\right>,\label{eq:efield}
\end{equation}
in cylindrical coordinates with $r=\sqrt{x^2+y^2}$. It has a radial profile described by a function $f(r)\in[0,1]$ with $f(0)=1$, and a time profile described by a function $w(t)\in[0,1]$. The time profile has a smooth rise and fall from/to zero (over time scales much larger than $2\pi/\omega_0$) surrounding a region where $w(t)=1$. 

Such an electric field may be produced near the lattice by a configuration of nano-scale electrodes. For example, a two-dipole configuration with orientations at $\qty{90}{\degree}$ from each other, and with oscillating voltages with a relative delay of $\pi/2\omega_0$ has a component that can be described by \eqref{eq:efield}. 

% For some illustrations we consider a one-parameter exponential radial profile $f(r)=\exp(-r/r_0)$. 
% For the results in \S\ref{sec:results}, we consider a more realistic situation, 
For the radial profile, we consider an electric potential that falls off as the modified Bessel function of the second kind $K_0(r/r_0)$, solution to the screened Poisson equation with cylindrical symmetry. This is equivalent to assuming that the electric field cannot travel outside the material and avoid screening, producing a fast fall-off. For that reason, the results presented here provide a conservative estimate of the strength of the magnetic effects of this field. Out-of-plane field propagation would increase the field value at longer distances and only strengthen the effects presented here. The electric field would present a radial profile $f(r)\propto K_1(r/r_0)$, which behaves as $1/r$ at short distances. To prevent artificially large effects that can't be realized experimentally, we clip the radial profile below a distance $r_\mathrm{min}$ such that
 \begin{equation}
 	f(r)=\begin{cases} 1, & r\leq r_\mathrm{min},\\ K_1(r/r_0)/K_1(r_\mathrm{min}/r_0),&r>r_\mathrm{min},\end{cases}
	\label{eq:besselprofile}
 \end{equation}
 where $r_\mathrm{min}$ would be around the size of the structures producing the electric field. 
%  These models are illustrated in Fig.~\ref{fig:profiles}.

%  \begin{figure}\centering\includegraphics[width=\linewidth]{profiles.pdf}\caption{Different radial profiles used throughout this work.}\label{fig:profiles}\end{figure}

The rest of the model is contingent upon the choice of materials one decides to consider. Here, we don't aim to simulate a real material. Rather, we present a mechanism by which the effect of such an electric field can make its way to impact the magnetic evolution. Whether this requires a single material or a more complex multi-layered structure is out of the scope of this work.

The electric field makes its impact on the lattice via its force density $\femv$. There is a long-standing controversy about the form of the force density exerted by electromagnetic fields on matter~\cite{barnettelectromagnetic2006,shevchenloelectromagnetic2011,friaselectromagnetic2012,anghinonimicroscopic2023}. Since our interest lies in illustrating the feasibility of an effect rather than its precise quantification, we don't address this directly, and simply describe the impact of the field on matter using the electric part of the Maxwell tensor,~\cite{LL-ECM}
\begin{equation}
	T_{ij}=E_iD_j-\frac12\delta_{ij} \bm E\cdot\bm D,
\end{equation}
where $\bm D=\epsilon_0\bm E+\bm P$ and $\bm P$ are the macroscopic displacement and polarization vectors, respectively. We restrict this work to linear isotropic materials, for which $\bm P\propto\bm E$ and, thus,  $\bm D=\epsilon\bm E$, where $\epsilon$ is the electric permittivity of the material. From this, we may write
\begin{equation}
	T_{ij}=\epsilon\left(E_iE_j-\frac12\delta_{ij} E^2\right),\label{eq:maxwelltensor}
\end{equation}
which may be used to obtain the force density.

The external field introduces a deformation on the material, described by $\bm u(t,\bm x)$, such that the position of the $i$-th particle in the lattice becomes $\bm x_i=\bm x_i^\mathrm{eq}+\bm u(t,\bm x_i^\mathrm{eq})$, where $\bm x_i^\mathrm{eq}$ is its equilibrium position. These deformations give rise to internal elastic forces, of components $\felv$. Considering an isotropic solid, the elastic force density is~\cite{landautheory1986}
\begin{equation}
	\felv = (\mu+\lambda)\bm\nabla (\bm\nabla\cdot\bm u) + \mu\Delta\bm u,
\end{equation}
where $\Delta=\bm\nabla^2$ and $\mu$ and $\lambda$ are first and second Lam\'e parameters ($\mu$ is also known as shear modulus).

Combining the effects of $\femv$ and $\felv$ we obtain the Navier--Cauchy equation in the presence of an external force $\femv$:
\begin{equation}
	\rho\ddot{\bm u} = (\mu+\lambda)\bm\nabla (\bm\nabla\cdot\bm u) + \mu\Delta\bm u + \femv.\label{eq:naviercauchy}
\end{equation}

As we show below, only the rotation and angular velocity fields
\begin{equation}
	\bm\phi=\bm\nabla\times\bm u\quad\text{and}\quad\bm\Omega=\dot{\bm\phi}=\bm\nabla\times\dot{\bm u}\label{eq:rotfields}
\end{equation}
play a role in the magnetic evolution of the lattice. If the electric field is mostly in-plane ($\varepsilon_z\ll 1$), the deformation field is constrained to the $x-y$ plane. Thus, we may write $\bm\phi=\phi\,\hat{\mathbf z}$ and $\bm\Omega=\Omega\,\hat{\mathbf z}$. In the appendix, we obtain the solution to \eqref{eq:naviercauchy} for $\phi$ in the steady state ($w(t)=1$) as
\begin{equation}
\phi(t,r,\theta)=\phi^s(r)\sin(2\theta-2\omega_0t) + \phi^c(r)\cos(2\theta-2\omega_0t),\label{eq:phi}
\end{equation}
where the radial dependence is captured by
\begin{subequations}
	\begin{align}
	\phi^s(r) & =\frac{\iota}{\mu}\left\{F(r) -\frac{\pi}{2}\left[ I_Y^\infty(kr)J_2(kr) + I_J^0(kr)Y_2(kr)\right]\right\},\\
	\phi^c(r) & = \frac{\iota}{\mu}\frac{\pi}{2}I_J^{\mathbb R_+}J_2(kr).
	\end{align}\label{eq:phi-2}
\end{subequations}
The above definitions contain the Bessel functions of the first ($J$) and second ($Y$) kind, as well as their integrals ($I_J$ and $I_Y$) over different intervals (represented by their superscripts, see Appendix). The radial profile of the field is contained in $F(r)$, and $k=2\omega_0/c_\mathrm t$, with $c_\mathrm t=\sqrt{\mu/\rho}$, the speed of transverse waves. The strength of the electric field plays a role via $\iota=\epsilon E_0^2/2$.
From \eqref{eq:phi}, a similar expression is found for the angular velocity field:
\begin{equation}
	\Omega(t,r,\theta)=\Omega^s(r)\sin(2\theta-2\omega_0t) + \Omega^c(r)\cos(2\theta-2\omega_0t),\label{eq:omega}
\end{equation}
with
\begin{equation}	
	\Omega^s(r)  = 2\omega_0\phi^c(r)\quad\text{and}\quad
	\Omega^c(r)  = -2\omega_0\phi^s(r)\,.
	\label{eq:omega-2}
\end{equation}

We now turn to the magnetic description of the lattice. We evolve the spin vector at each lattice site, $\bm S_i$, via the Landau--Lifshitz equation subject to an effective field
\begin{equation}
	\bm H^\mathrm{eff}_i=-\frac{1}{\gamma\hbar} \frac{\partial\mathcal H}{\partial\bm S_i}\,.\label{eq:heff:def}
\end{equation}
The magnetic Hamiltonian,
\begin{multline}
	\mathcal H=
		-\frac J2\sum_{i,j}\bm S_i\cdot\bm S_j
		-\frac D2\sum_i\left(\mathbf n_i\cdot \bm S_i\right)^2\\
		-\gamma\hbar\sum_i\bm S_i\cdot\bm H
		-\hbar\sum_i\bm S_i\cdot\bm\Omega_i\,,\label{eq:hamiltonian}
\end{multline}
contains nearest-neighbor exchange interactions, with energy $J$, uniaxial magnetic anisotropy, with energy $D$, an external magnetic field $\bm H$, and the non-inertial term produced by the local angular velocity of the lattice, $\bm\Omega$. The exchange sum extends to nearest neighbors only.

The time-dependent lattice deformations introduce a difference between the local rest frame at each lattice site and the lab frame. We write the lattice-frame equation for the $i$-th spin as~\cite{EC-JS-EPL2026}
\begin{equation}
	\frac{\upd{\bm S_i}}{\upd t}=\gamma{\bm S_i}\times(\bm H^\mathrm{eff}_i-\bm\Omega_i/\gamma)-\alpha\gamma\bm S_i\times(\bm S_i\times\bm H^\mathrm{eff}_i)\,,\label{eq:ll}
\end{equation}
where the effective field can be obtained from \eqref{eq:heff:def} and \eqref{eq:hamiltonian} as
\begin{equation}
	\bm H^\mathrm{eff}_i=\bm H + \frac{1}{\gamma\hbar}\left(J\sum_j \bm S_j+D\left(\bm S_i\cdot \mathbf n_i\right)\mathbf n_i+\hbar\bm\Omega_i\right)\,.\label{eq:heff}
\end{equation}
All subscripted quantities are evaluated at $\bm x=\bm x_i^\mathrm{eq}$.

Notice that the effects of the local rotations are twofold. First, they enter $\bm H^\mathrm{eff}_i$ directly via $\bm\Omega$, although this term is cancelled in the gyroscopic term of the Landau--Lifshitz equation. Second, they enter through the anisotropy term as follows: the anisotropy axis is \emph{attached} to the local frame of the lattice, where it has a constant value $\tilde{\mathbf n}$. In the lattice frame, this vector will rotate by a different angle $\phi_i$ about the $z$ axis at each lattice site. Thus, $\mathbf n_i=R_z[\phi_i]\tilde{\mathbf n}$, introducing the effect of the deformations both in the gyroscopic and damping terms.

\section{The deformation fields}
As is clear from \eqref{eq:ll} and \eqref{eq:heff}, only $\bm\phi=\bm\nabla\times\bm u$ and $\bm\Omega=\bm\nabla\times\dot{\bm u}$ are necessary to understand the impact of the deformations on the magnetic dynamics. 
In the appendix, we partially solve the Navier--Cauchy equation \eqref{eq:naviercauchy} for $\bm\phi$ and $\bm\Omega$ with a finiteness condition at $r=0$ and the Sommerfeld radiation condition at $r\to\infty$. %The latter condition makes the solution valid only as long as the radial extension of the electric field is much smaller than the lattice size.

\begin{figure}\centering\includegraphics[width=\linewidth]{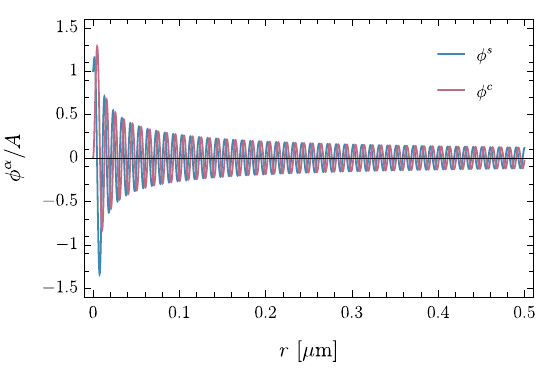}\caption{Rescaled radial functions $\phi^s$ and $\phi^c$ for an external electric field with the radial profile from \eqref{eq:besselprofile} with $r_\mathrm{min}=\qty{10}{nm}$, $r_0=\qty{1}{\micro m}$, a speed of sound $c_\mathrm t=\qty{3.16}{km/s}$, and angular frequency $\omega_0=\qty{1}{THz}$ ($f_0\approx\qty{160}{GHz}$).}\label{fig:radial}\end{figure}

\begin{figure}\centering\includegraphics[width=\linewidth]{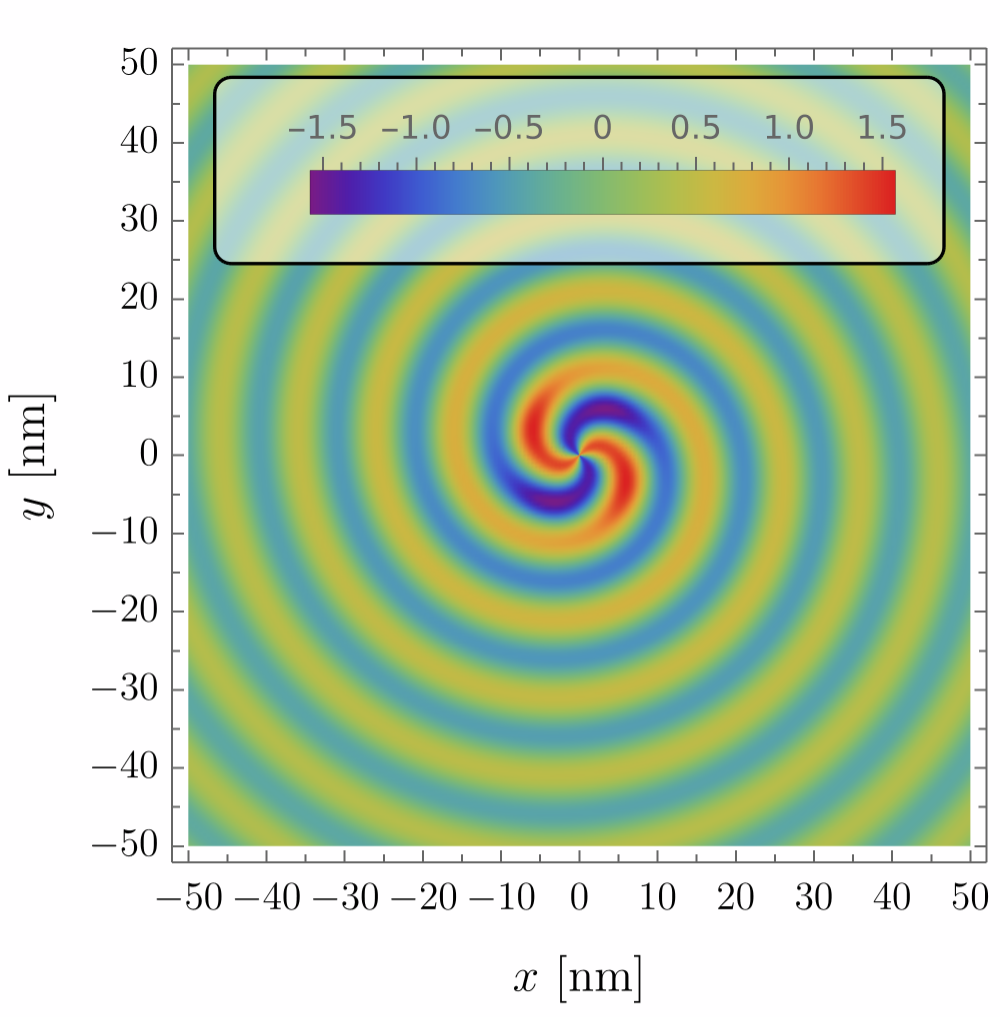}\caption{Rescaled rotation field $\phi$ at $t=0$, for the same parameter choices as in Fig.~\ref{fig:radial}. The pattern shown here rotates counter-clockwise as time passes.}\label{fig:full}\end{figure}

In this section, we provide a description of the rotation field $\phi$ from \eqref{eq:phi} and, in particular, its radial dependence, described in \eqref{eq:phi-2}. These quantities are proportional to the dimensionless ratio $A=\iota/\mu$, which depends both on the field and the material properties. 
We illustrate the values of $\phi^\alpha(r)/A$ (with $\alpha=c,s$) and $\phi/A$ in Fig.~\ref{fig:radial} and Fig.~\ref{fig:full}. To do so,  we choose an electric field with radial profile \eqref{eq:besselprofile}, with $r_\mathrm{min}=\qty{10}{nm}$, $r_0=\qty{1}{\micro m}$, $c_\mathrm t\approx\qty{3}{km/s}$, and $\omega_0=\qty{1}{THz}$.

\begin{figure}\centering\includegraphics[width=\linewidth]{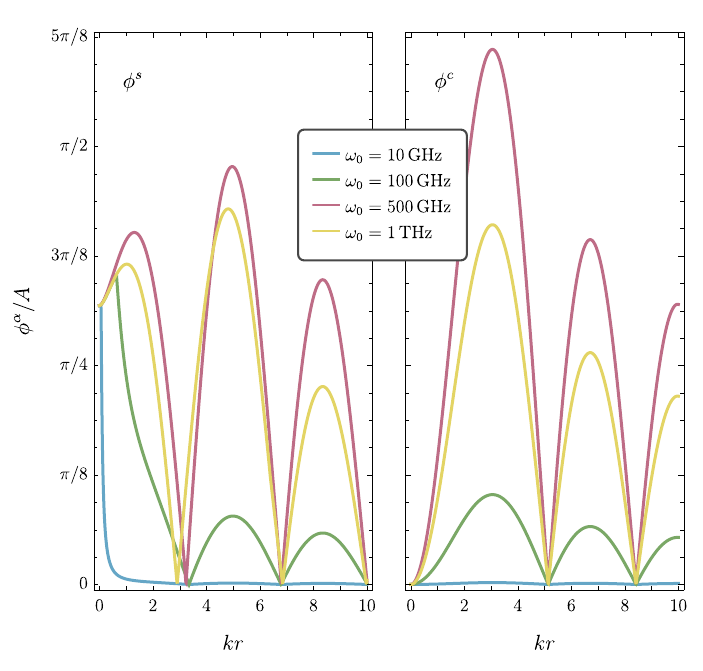}\caption{Rescaled radial functions $\phi^s/A$ (left) and $\phi^c/A$ (right) for different frequencies $\omega_0$ such that $kr_0=2\omega_0r_0/c_\mathrm{t}$ takes values $0.1$, $1$, $10$ and $100$, with $r_0=\qty{1}{\mu m}$ and $c_\mathrm t=\qty{3}{km/s}$. This corresponds to angular frequencies $\omega_0$ between \qty{0.15}{GHz} and \qty{0.15}{THz} ($f_0$ between \qty{24}{MHz} and \qty{24}{GHz}).}\label{fig:phi-kr0}\end{figure}

While the frequency of the field doesn't impact the value of $A$, it has an impact on the radial profile of the solution, in two ways. First, the Bessel functions exhibit oscillations on the radial scale $k^{-1}$. Second, the amplitude of such oscillations is determined by the relation between $k$ and $r_0$, as follows. In the high frequency limit, where $kr_0\gg1$, the radial profile of the field barely changes throughout the oscillation scale $k^{-1}$. Thus, near the center of the field, the amplitude of the oscillations is determined by the $1/\sqrt{kr}$ decay of the Bessel functions. For frequencies such that $kr_0\sim1$ or lower, the radial profile changes significantly through the oscillation scale, so the amplitude of the oscillations is further damped by $F(r)$. This sets a frequency threshold	$\omega_\mathrm{th}={c_\mathrm t}/{2r_0}$ below which we expect the lattice to remain largely undisturbed by the electric field.
Moreover, the amplitude of $\phi^\alpha/A$ is maximum in the high-frequency limit $kr_0\gg1$. This is illustrated in Fig.~\ref{fig:phi-kr0}.
There, it can be seen that at low frequencies, $\phi^\alpha$ decreases very rapidly. As the frequency increases to a few hundreds of gigahertz, the maximum values of these functions reach slightly below $5\pi/8$. At terahertz frequencies and above, the frequency dependence in the maximum values disappears, and all functions are roughly equal to the yellow curve (\qty{1}{THz}) shown in the figure. All in all,  $\phi^\alpha/A$ stays roughly at and below $\pi/2$ in absolute value, so $\phi_i\lesssim A\pi/2$. Rotations larger than $\pi/2$ would be unphysical within the elastic theory, so we constrain $A$ to be approximately less than one. Throughout this work, we choose the more conservative estimate of $A=0.1$.

The angular velocity field $\bm\Omega$ presents the same spatial structure as $\bm\phi$, with the exception of an additional term $\omega_0$, as seen in \eqref{eq:omega-2}. Thus, the amplitude of the oscillations in $\bm\Omega$ increases with $\omega_0$ beyond the maximum amplitude reached by $\phi$ for $kr_0\gg1$. From \eqref{eq:heff} and \eqref{eq:omega-2}, one can see that the $\bm\Omega$ term will have a non-negligible effect on the effective field when $A\hbar\omega_0\gtrsim D$. Nevertheless, since this term is cancelled in the gyroscopic part of the Landau--Lifshitz equation, its impact on the magnetic evolution will only be effective at a higher frequency, when $\alpha A\hbar\omega_0\gtrsim D$.

\section{Frequency regimes}
The effect of the external field on the magnetic evolution of the system depends on its ability to absorb energy from the field efficiently. The rich structure of the equations allows for a complex response across the frequency spectrum. Here, we exploit this structure to differentiate several regimes which are worth separating in our analysis.

First, as described above, frequencies below $\omega_\mathrm{th}$ are unlikely to produce any sizeable disturbance to the magnetic dynamics unless $A\gg1$, which we don't consider here.

Beyond that, the external field produces oscillations on the lattice with frequency $2\omega_0$, as seen in \eqref{eq:phi} and \eqref{eq:omega}. 
In a uniform magnetic lattice without external field, the Landau--Lifshitz equation may be linearized around its equilibrium. There, it can be seen that oscillations occur with angular frequency $\omega_\mathrm{mag}=D/\hbar$. Thus, we expect the energy transfer from the electric field to the magnetic lattice to magnify for $\omega_0\sim\omega_\mathrm{mag}/2$.

At low frequencies, the structure of the Landau--Lifshitz equation and the small value of the damping constant restrict the effect of the field to the gyroscopic term. In that term, the $\bm\Omega$ field is not present, so all effects are produced by the efficient coupling to the rotation field $\bm\phi$. As the frequency increases, the smallness of $\alpha$ can be compensated by the large value of $\omega_0$, to which $\bm\Omega$ is proportional. This behavior is expected at frequencies larger than $\omega_{\bm\Omega}=D/\alpha A\hbar$. 

Finally, the validity of the solution to the Navier--Cauchy equation breaks down for mechanical waves with wavelength $\lambda\lesssim 2\pi/k_D$, where $k_D\sim\pi c_\mathrm{t}/a$ is the Debye wavenumber, which corresponds to an angular frequency $\omega_D=\pi c_\mathrm{t}/a$.

To simplify our analysis, we fix several of the physical parameters of the problem and focus our attention on variations produced by different values of the frequency of the external field. As mentioned before, it is not our goal to find a material, or combination of materials, for which to provide a realistic simulation of this effect, but rather, to illustrate its feasibility. We choose the material properties typical of magnetic oxides, as $\rho=\qty{5e3}{kg/m^3}$, $\mu=\qty{50}{GPa}$, $c_\mathrm{t}\approx\qty{3.2e3}{m/s}$, and $\omega_D=\qty{50}{THz}$.

\section{Results}\label{sec:results}
We now present the results of our simulations as follows. First, we study the size of the magnetic deviations from the equilibrium produced by electric fields at different frequencies. Second, for those regions in which we identify a significant impact, we take a closer look at the magnetic dynamics and the occurrence of spin flips. Third, we analyze the impact that an external magnetic field has in aiding those reversals.  Moreover, we split the analysis in two frequency regimes: \textit{(i)} the gyroscopic regime with $\omega_0\sim\omega_\mathrm{mag}/2$, which excites the spins via the coupling with the time-varying anisotropy direction in the gyroscopic term of the LL equation; and \textit{(ii)} the damping regime, which is present at higher frequencies ($\omega_0\gtrsim\omega_{\bm\Omega}$), and driven by the angular velocity $\bm\Omega$ in the damping term of the LL equation. The first mechanism is enhanced in high-anisotropy and low-damping materials, while the second benefits from low anisotropy and high damping. 

If the external electric field has a vanishingly small $z$ component, the rotation and angular velocity fields are directed along the $z$ axis. Due to the structure of the Landau--Lifshitz equations, a magnetic uniaxial anisotropy along the $z$ axis would make the initial state insensitive to the field. Thus, we consider anisotropies directed along different directions in the $x-z$ plane, at an angle $\theta$ from the $+z$ direction. After fixing the anisotropy axis, there are two stable equilibria for a uniform lattice (for zero or small fields). We will study the possible reversal from one to the other by computing the magnetic evolution of the spins in the lattice according to \eqref{eq:ll}. 
For these simulations aiming to explore many frequency values (as well as varying other parameters), we choose a small lattice of $N\sim10^4$ spins. 

We quantify the ability of the electric field to disturb the lattice by calculating the time-dependent difference between the average spin and its equilibrium value. To do so, we define $\bm\mu(t)=\sum_{i=1}^N\bm S_i(t)/N$,  $\bm\mu_0=\bm\mu(t=0)$,
% and $\bm\mu_\shortparallel=\bm\mu-(\bm\mu\cdot\hat{\mathbf z})\,\hat{\mathbf z}$, 
and use 
% \begin{subequations}
% \begin{equation}
% 	\Delta_\shortparallel=\frac{1}{t_f-t_i}\int_{t_i}^{t_f}|\bm\mu_\shortparallel(t)-\bm\mu_{\shortparallel,0}|^2 \upd t
% \end{equation}
% and 
\begin{equation}
	\Delta=\frac{1}{t_f-t_i}\int_{t_i}^{t_f} |\mu_z(t)-\mu_{0,z}|^2 \upd t
\end{equation}
% \end{subequations}
which captures the average impact that the field has on the component of the spin perpendicular to the lattice. 

Throughout this section, we present results for the Bessel radial profile \eqref{eq:besselprofile} with $r_\mathrm{min}\approx\qty{10}{nm}$ with $A=0.1$ and $r_0=\qty{1}{\micro m}$.

\subsection{Gyroscopic regime}

Here, we analyze the behavior of the lattice for frequencies around $\omega_\mathrm{mag}/2$. We consider magnetic anisotropy energies of $D=\qtyrange{0.005}{0.5}{meV}$, corresponding to angular frequencies $\omega_\mathrm{mag}\approx\qtyrange{8}{800}{GHz}$. In this regime, the only effect of $\alpha$ is to damp the oscillations of the spins around their equilibrium point, which results in lowering the magnitude of the absorption peaks and making it more difficult to potentially achieve a reversal. We keep it at a fixed value of $\alpha=0.001$. We also probe exchange energies $J=\qtyrange{1}{30}{meV}$.

\begin{figure}\centering
	\includegraphics[width=\linewidth]{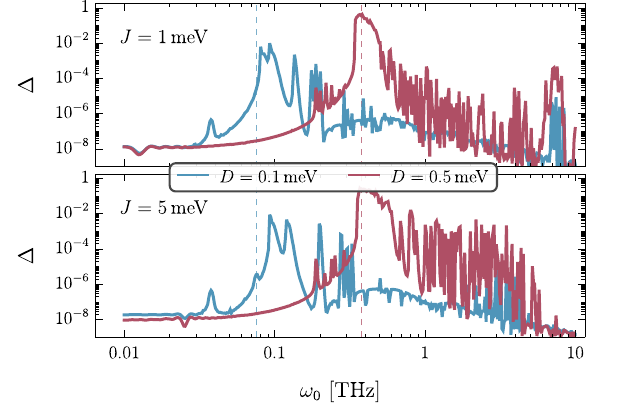}\caption{Time-averaged square deviation of the $z$-component of the average lattice spin $\bm\mu$ as a function of the frequency of the external field, for $D=\qty{0.1}{meV}$ (blue), $D=\qty{0.5}{meV}$ (red), $J=\qty{1}{meV}$ (top) and $J=\qty{5}{meV}$ (bottom). The vertical lines indicate the values $\omega_0=\omega_\mathrm{mag}/2\approx\qtylist{76;380}{GHz}$ ($f_0\approx\qtylist{12;60}{GHz}.$).}\label{fig:absorption:gyro}\end{figure}

In Fig.~\ref{fig:absorption:gyro}, we illustrate the ability of the electric field to disturb the lattice in this regime, for anisotropy energies $D=\qtylist{0.1;0.5}{meV}$ (corresponding to $\omega_\mathrm{mag}\approx\qtylist{150;760}{GHz}$), and exchange energies $J=\qtylist{1;5}{meV}$. The main resonances are observed at frequencies $\omega_0\sim\omega_\mathrm{mag}/2$, and a complex structure is seen at larger frequencies.

\begin{figure}\centering
	\includegraphics[width=\linewidth]{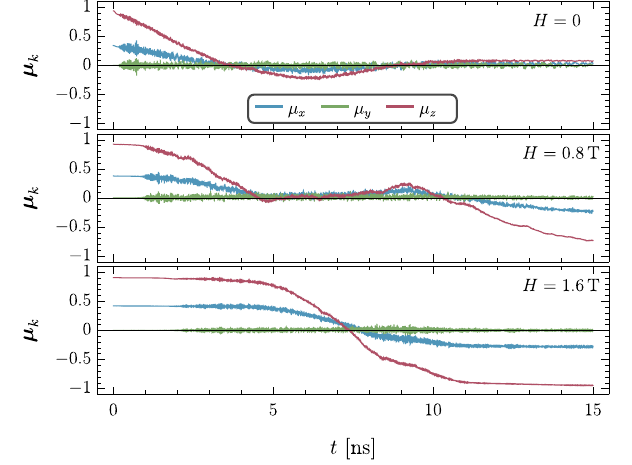}\caption{Evolution of average spin in a lattice with $D=\qty{0.5}{meV}$, $J=\qty{5}{meV}$, under a \qty{10}{ns} duration electric field with $\omega_0=\qty{400}{GHz}$ ($f_0=\qty{64}{GHz}$) and magnetic fields $H=\qtylist{0;0.8;1.6}{T}$.}\label{fig:evolution-gyro}\end{figure}

\begin{figure}\centering
	\includegraphics[width=0.9\linewidth]{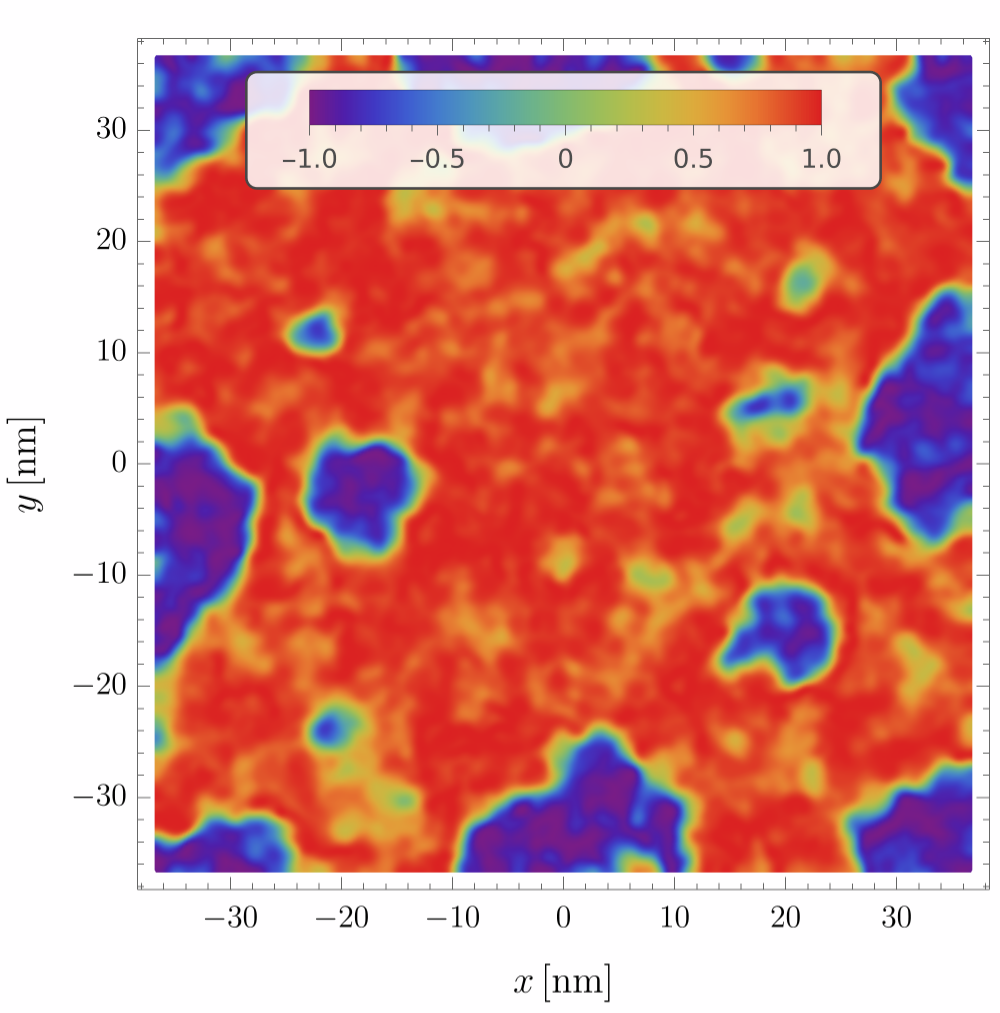}\caption{Intermediate formation of magnetic domains during the application of the external electric field for $D=\qty{0.5}{meV}$, $J=\qty{5}{meV}$, $\alpha=0.001$, $\omega_0=\qty{400}{GHz}$ ($f_0=\qty{64}{GHz}$), and $H=0$, for a lattice of approximately $10^5$ spins. The colors correspond to the values of the $z$-component of the spins.}\label{fig:gyro:lattice}\end{figure}

For a specific example of the impact of the field on the lattice, we choose $D=\qty{0.5}{meV}$, $J=\qty{5}{meV}$, and a frequency $\omega_0=\qty{400}{GHz}$. The time evolution of the average spin is shown in Fig.~\ref{fig:evolution-gyro} for a larger lattice of $N=10^5$ spins. The external electric field is applied for about \qty{10}{ns} and, in the absence of an external magnetic field, it leads to the chaotic creation of magnetic domains (see Fig.\ref{fig:gyro:lattice}) that result in a near-zero magnetization. After the external electric field is switched off, the spins commence a relaxation process on a time scale dictated by $\alpha$, into either of the two equilibria aligned with the anisotropy axis. The chaotic nature of the domain formation makes it unpredictable which of the two equilibria the spins will relax into.

In order to make reversals consistent, we may add an external magnetic field pointing in the $-z$ direction. If the magnetic field magnitude is small, this creates a metastable equilibrium with positive $z$ component, at an angle $\theta_0$ from the $+z$ axis, which we choose as the initial state. Examples of such reversals are shown in Fig.~\ref{fig:evolution-gyro} (middle and bottom).

	\subsection{Damping regime}
	We now proceed to analyze the behavior at higher frequencies. In this regime, the damping term grows indefinitely with $\omega_0$ without increasing $\phi$ and, thus, without surpassing the regime of validity of the elastic approximation. With a judicious choice of the magnetic parameters, we shall see that the impact on the magnetic evolution of the lattice can be much greater than in the gyroscopic regime.

	The effect of the damping coefficient is to decrease the frequency $\omega_0$ at which the effects of the damping term dominate; thus, large $\alpha$ should favor spin reversals. The anisotropy energy contributes to aligning the spins with their initial equilibrium, so lower anisotropies make it easier to leave that equilibrium. As expected, higher frequencies contribute as well to increasing the effect. For illustration, we choose $\alpha=0.2$, $D=\qty{0.005}{meV}$, $J\approx\qty{7.2}{meV}$ and a frequency $\omega_0=\omega_D$. 
	
\begin{figure}\centering
	\includegraphics[width=\linewidth]{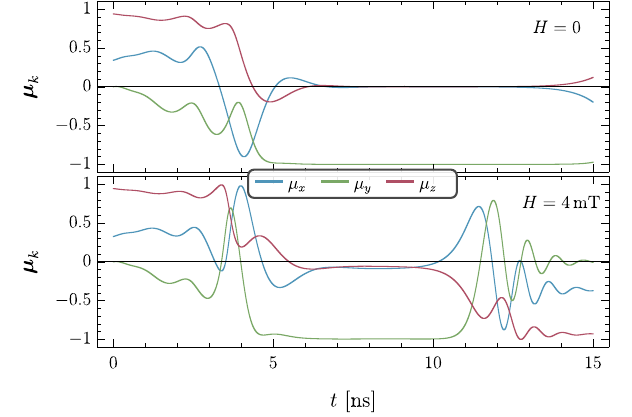}\caption{Evolution of average spin in a lattice with $D=\qty{0.005}{meV}$, $J=\qty{7.22}{meV}$, under a \qty{10}{ns} duration electric field with $\omega_0=\omega_D=\qty{50}{THz}$ ($f_0\approx\qty{8}{THz}$) and magnetic fields $H=\qtylist{0;4}{mT}$.}\label{fig:evolution-damp}\end{figure}

\begin{figure}\centering
	\includegraphics[width=0.8\linewidth]{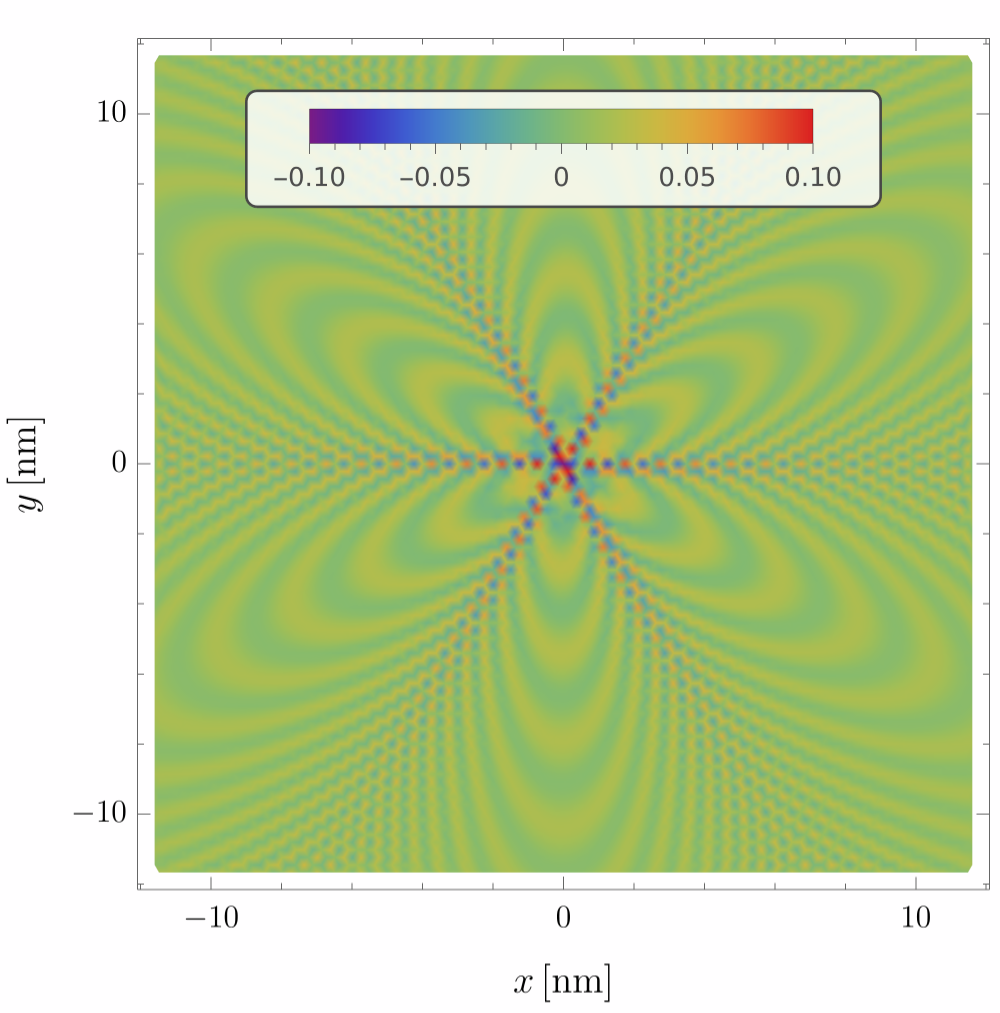}\caption{Intermediate formation of spin waves during the application of the external electric field for $D=\qty{0.005}{meV}$, $J=\qty{7.22}{meV}$, $\alpha=0.2$, $\omega_0=\qty{50}{THz}$ ($f_0\approx\qty{8}{THz}$), and $H=0$, for a lattice of approximately $10^5$ spins. The colors correspond to the values of the $z$-component of the spins.}\label{fig:damp:lattice}\end{figure}

	Upon application of the external electric field, the $z$-component of the spin goes to zero but, unlike in the gyroscopic regime, there is no formation of magnetic domains. Rather, the spins are aligned along the $y$ axis quite uniformly, as can be seen in Fig.~\ref{fig:evolution-damp} . Small deviations around this direction appear in the form of outward-propagating spin waves (see Fig.~\ref{fig:damp:lattice}).

\section{Discussion}
We have proposed an electro-mechanical mechanism by which a rotating electric field induces local rotations on the film, which ultimately have an effect on the magnetization via local frame-dependent effects.

Choosing a simple configuration of the rotating electric field, we proceeded to solve the Navier--Cauchy equations of elasticity under the external force produced by the electric field. We constrained the strength of the external field and its frequency to remain in the small-angle and sub-Debye regimes, which allows us to use the elastic theory. Using the rotation and angular velocity fields derived in this framework, we solved the Landau--Lifshitz equation containing the local frame-dependent effects introduced by these rotations. 

For numerical analysis, we chose typical parameters of weakly conducting magnetic oxides. Our microscopic spin model includes exchange interaction, magnetic anisotropy, and phenomenological damping. The inclusion of the dipole-dipole interaction greatly complicates the problem without yielding qualitative differences; we ignore it, assuming that it is lower in strength than other interactions. 

We defined two regimes: \emph{(i)} one dominated by the gyroscopic term of the Landau--Lifshitz equation via local changes to the anisotropy direction, which is most relevant in materials with low damping and high anisotropy, and \emph{(ii)} another dominated by the relaxation term, via the non-inertial effects produced by the local rotations, which is most relevant in materials with high damping and low anisotropy. In both regimes, we have illustrated the effects of these phenomena and demonstrated the existence of magnetization reversals.

In the first regime, magnetization reversals are favored by a strong anisotropy and weak damping. This enhances the impact of the local changes in the anisotropy direction, but allows the gyroscopic excitations to build up without being damped away. Nevertheless, some damping is necessary for the reversal to complete via a relaxation to a stable equilibrium on a time scale dictated by $\alpha$.

In the second regime, reversals are favored by low anisotropy and high damping. They are driven by the Barnett effect, the essence of which is the relaxation of the spin toward the direction of the effective field formed by mechanical rotations and magnetic interactions in the non-inertial coordinate frame coupled to the crystal lattice. Thus, damping does not just play the role of relaxing to a final equilibrium, but it is at the core of the phenomenon that brings the spins out of their initial equilibrium.

We have demonstrated that the effect of the rotating electric field on the magnetization is significant when $\omega_0\gtrsim\omega_\mathrm{mag}/2=D/2\hbar$ (gyroscopic regime) or $\alpha A \hbar\omega_0 \gtrsim D$ (damping regime), where $\alpha < 1$ is a damping factor, $A < 1$ is the ratio of the electromagnetic and elastic energy, $\omega_0 = 2\pi f_0$ is the angular frequency of the electric field, and $D$ is the energy of the magnetic anisotropy. %To satisfy this condition, the frequency $f_0$ must be sufficiently high. Magnetization reversal has been observed at  $f_0 = \omega_0/2\pi$ of a few tens  of gigahertz. 

The frequency range considered in this work, extending from several tens gigahertz to several terahertz, lies within the capabilities of existing gigahertz, sub-terahertz and terahertz technologies. A pair of orthogonal electrodes driven in quadrature, as in Fig.~\ref{setup}, or any equivalent near-field structure capable of producing a rotating in-plane electric field, would be plausible at the bottom of that frequency range. For higher frequencies the rotating electric field could be generated by electronic frequency multipliers, photoconductive antennas, photomixers, or ultrafast optical techniques, for which Fig. 1 should be regarded as an idealized realization of field configuration rather than a specific engineering design.

% Such frequencies are technologically accessible with sub-terahertz sources, photoconductive antennas, or frequency-multiplied microwave electronics, although producing a well-controlled rotating near field at nanoscale electrodes may be demanding. The electrode geometry in Fig. 1 should be regarded as an idealized local-field model rather than a ready-made microwave circuit.

Experimental confirmation of the principal possibility of the proposed effect, and its further exploration, will establish its potential for  practical applications. 

\acknowledgments
This work has been supported by Grants No. FA9550-24-1-0090 and FA9550-24-1-0290, funded by the U.S. Air Force Office of Scientific Research.

\appendix
\section{Solution to the Navier--Cauchy equation}
Below, we present the solution of the Navier--Cauchy equation \eqref{eq:naviercauchy} with a force density derived from the stress tensor \eqref{eq:maxwelltensor}, for the electric field \eqref{eq:efield}, in a steady situation with $w(t)=1$. We present only a partial solution: instead of solving for the deformation field $\bm u$, we solve only for the rotation fields \eqref{eq:rotfields}, as this is the only input required to solve the Landau--Lifshitz equation \eqref{eq:ll}.

The force density may be obtained from the stress tensor as
\begin{equation}
	f^\mathrm{em}_i=\partial_jT_{ij}-\frac{\partial g_i}{\partial t},
\end{equation}
where $\bm g=\bm E\times\bm H/c^2$ is the electromagnetic momentum density, which we neglect due to the smallness of the magnetic fields produced by the time variations in $\bm E$.

Using \eqref{eq:efield} and \eqref{eq:maxwelltensor},
\begin{equation}
	T^\mathrm{em}_{ij}=\epsilon E_0^2f^2(r)\left(\mathcal A_{ij}-\frac12\delta_{ij}\right),
\end{equation}
where 
\begin{equation}
	\mathcal A_{ij}(t)=\begin{pmatrix}
		\cos^2(\omega_0 t) & \sin(\omega_0 t) \cos(\omega_0 t)& 0\\
		\cos(\omega_0 t) \sin(\omega_0 t) & \sin^2(\omega_0 t)	 & 0\\
		0&0&0
		\end{pmatrix}.\label{eq:Aij}
\end{equation}
at $z=0$. Then, the third component of the electromagnetic force vanishes, and $u_z$ is constant. The problem is thus trivially reduced to the $x-y$ plane. From now on, we introduce indices $a,b,c,\dots$ for 2D summations, and leave $i,j,k,\dots$ for 3D summations.

To solve the 2D equation of motion, we use a two-dimensional Helmholtz decomposition for $\bm u$ and $\femv$ without harmonic component:
\begin{subequations}
\begin{align}
	u_a&=\partial_a\varphi+\varepsilon_{ab}\partial_b\psi,\label{eq:helmholtz:u}\\
	f^\mathrm{em}_a&=\partial_a\chi+\varepsilon_{ab}\partial_b\eta,\label{eq:helmholtz:fem}
\end{align}
\end{subequations}
where $\varepsilon_{ab}$ are the components of the rank 2 antisymmetric tensor,
\begin{equation}
	[\varepsilon_{ab}] = \begin{bmatrix}0&1\\-1&0\end{bmatrix}.
\end{equation}
Introducing this in \eqref{eq:naviercauchy}, two equivalent equations are found for the scalars $\varphi$ and $\psi$:
\begin{subequations}
\begin{align}
\varrho\ddot\varphi&=(\lambda+2\mu)\partial_a\partial_a\varphi+\chi,\label{eq:eom2d:phi}\\
\varrho\ddot\psi&=\mu\partial_a\partial_a\psi+\eta.\label{eq:eom2d:psi}
\end{align}
\end{subequations}

We may now see that $\bm\phi=\bm\nabla\times\bm u=(\partial_x u_y-\partial_y u_x)\hat{\mathbf z}$, which allows to rewrite
\begin{equation}
	\phi=\bm \phi\cdot\hat{\mathbf z}=\partial_x u_y-\partial_y u_x=-\Delta\psi,
\end{equation}
where $\Delta=\partial_a\partial_a$. Thus, only \eqref{eq:eom2d:psi} needs to be solved to account for the effects of the electric field on the Landau--Lifshitz equation. To do this, we first invert \eqref{eq:helmholtz:fem} to find $\eta$, which satisfies the Poisson equation
\begin{equation}
	\partial_a\partial_a\eta=\varepsilon_{ab}\partial_b f_a^\mathrm{em}.
\end{equation}
Using the Green's function for the Laplacian in 2D, $G(\bm x,\bm x')=\log\left|\bm x-\bm x'\right|/2\pi$, and the relation $f^\mathrm{em}_a=\partial_c T^{\mathrm{em}}_{ac}$ we write
\begin{equation}
	\eta(\bm x)=\varepsilon_{ab}\int_{\mathbb R^2}G(\bm x,\bm x')\partial_b'\partial_c' T_{ac}^\mathrm{em}(\bm x') \mathrm d^2\bm x'.
\end{equation}
The integrand may be rewritten as
\begin{multline}
	G(\bm x,\bm x')\partial_b'\partial_c' T_{ac}^\mathrm{em}(\bm x')= \partial'_b\left[G(\bm x,\bm x')f_a^\mathrm{em}(\bm x')\right.\\
	\left.-\partial'_c\left(G(\bm x,\bm x')T_{ac}^\mathrm{em}(\bm x')\right)\right] +T_{ac}^\mathrm{em}(\bm x')\partial'_b\partial'_c G(\bm x,\bm x').
\end{multline}
In a large lattice, the exponential falloff of $T_{ab}$ guarantees that the boundary terms vanish. The derivative of the Green's function is 
\begin{equation}
	\partial'_b\partial'_c G(\bm x,\bm x')=\frac{1}{2\pi}\left[\frac{\delta_{bc}}{\left|\bm x-\bm x'\right|^2}-2\frac{(x_b'-x_b)(x_c'-x_c)}{\left|\bm x-\bm x'\right|^4}\right].
\end{equation}
Due to symmetry considerations, the first term does not contribute to $\eta$. A similar argument applies to the $\delta_{ac}$ term inside $T_{ac}^\mathrm{em}$. In the end, we can write
\begin{equation}
	\eta(\bm x)=\frac{2\iota}{\pi} \mathcal A_{ac}\int_{\mathbb R^2}
	 \varepsilon_{ba}\frac{(x_b'-x_b)(x_c'-x_c)}{\left|\bm x-\bm x'\right|^4}f^2(r')\mathrm d^2\bm x',
\end{equation}
where $\iota=\epsilon E_0^2/2$.
Defining the integrals
\begin{equation}
	I_{ab}(\bm x)=\int_{\mathbb R^2}
	 \frac{(x_a'-x_a)(x_b'-x_b)}{\left|\bm x-\bm x'\right|^4}f^2(r')\mathrm d^2\bm x',
\end{equation}
we may rewrite
\begin{multline}
	\eta(t,\bm x)=\frac{2\iota}{\pi} \mathcal A_{ac}\varepsilon_{ba} I_{bc}
	=\frac{2\iota}{\pi}\tr(\varepsilon\mathcal A I)
	\\=\frac{2\iota}{\pi}\left(\frac{I_{xx}-I_{yy}}{2}\sin2\omega_0t - I_{xy}\cos2\omega_0t\right).
\end{multline}
These integrals may be further simplified in polar coordinates, which allows to write
\begin{subequations}
\begin{align}
	\frac{I_{xx}-I_{yy}}{2}&=\frac{\pi\cos2\theta}{r^2}\int_0^r f^2(r')r' dr'\\
	I_{xy}&=\frac{\pi\sin2\theta}{r^2}\int_0^r f^2(r')r' dr'
\end{align}
\end{subequations}
Defining
\begin{equation}
	F(r)=\frac{2}{r^2}\int_0^r f^2(r')r' dr',
\end{equation}
we may finally write
\begin{multline}
	\eta(t,\bm x)=\iota F(r)\left(\cos2\theta\sin2\omega_0t - \sin2\theta\cos2\omega_0t\right)
	\\=\iota F(r)\sin(2\theta-2\omega_0t).
\end{multline}

We may now solve \eqref{eq:eom2d:psi}, rewriting it as
\begin{equation}
	(\partial_t^2-c_\mathrm t^2\Delta)\psi(t,r,\theta)=\eta(t,r,\theta)/\rho,
\end{equation}
where $c_\mathrm t=\sqrt{\mu/\rho}$ is the speed of transversal waves in the material, and $\Delta=\partial_r^2+r^{-1}\partial_r+r^{-2}\partial_\theta^2$. To do so, we consider the complexified functions $\tilde\psi$ and $\tilde\eta$, such that $\eta=\Im\{\tilde\eta\}$. Since the differential operator is real, $\psi=\Im\{\tilde\psi\}$, where $\tilde\psi$ solves
\begin{equation}
	(\partial_t^2-c_\mathrm t^2\Delta)\tilde\psi(t,r,\theta)=\frac{\tilde\eta(t,r,\theta)}{\rho}=\frac{\iota F(r)}{\rho}e^{i(2\theta-2\omega_0t)}.
\end{equation}
We can now successively expand $\tilde\psi$ in terms of the eigenfunctions of $\partial_t^2$ and $\partial_\theta^2$. We disregard the terms that constitute the homogeneous part of the solution, since they correspond to an initial state that is not driven by the external force. By doing so, we can write
\begin{equation}
	\tilde\psi(t,r,\theta)=\tilde\Psi(r) e^{i(2\theta-2\omega_0t)}.
\end{equation}
Then, $\partial_t^2\psi=-4\omega_0^2\psi$ and $\partial_\theta^2\psi=-4\psi$, which allows to obtain an equation for the radial part
\begin{equation}
	\tilde\Psi''(r)+\frac1r\tilde\Psi'(r)+\left(k^2-\frac4{r^2}\right)\tilde\Psi(r)=-\frac{\iota}{\mu} F(r),
\end{equation}
where $k=2\omega_0/c_\mathrm t$. 

We define $x=kr$ and $\bar\Psi(x)=\tilde\Psi(x/k)$, and find the equivalent equation
\begin{equation}
	\bar\Psi''(x)+\frac1x\bar\Psi'(x)+\left(1-\frac4{x^2}\right)\bar\Psi(x)=g(x),\label{eq:r:dimless}
\end{equation}
where $g(x)=-\iota F(x/k)/\mu k^2$.
This is the Bessel equation of order 2. We write its solution in terms of the Hankel functions
\begin{equation}
	H_\nu^\pm(x)=J_\nu(x)\pm i Y_\nu(x),\label{eq:hankel2JY}
\end{equation}
as a sum of a homogeneous solution $\bar\Psi_h$ and a particular solution $\bar\Psi_p$. The homogeneous solution is simply
\begin{equation}
	\bar\Psi_h(x)=AH_2^+(x) + BH_2^-(x).
\end{equation}
The particular solution is built using the method of variation of parameters as
\begin{multline}
	\bar\Psi_p(x)=
	-
	H_2^+(x)\int_a^x\frac{H_2^-(x')g(x')}{W(x')}dx'
	\\+
	H_2^-(x)\int_b^x\frac{H_2^+(x')g(x')}{W(x')}dx',
\end{multline}
where $W(x)=4/i\pi x$ is the Wronskian for $H_2^+$ and $H_2^-$. In order to constrain $A$ and $B$ we impose \emph{(i)} an outgoing radiation solution as $x\to\infty$, and \emph{(ii)} a finite solution at $x=0$. The asymptotic behavior of the Hankel functions is $H_2^\pm\sim e^{\pm ix}/\sqrt x$. Thus, upon combination with the harmonic term $e^{-2i\omega_0t}$, the $H_2^-$ term describes incoming radiation, not produced by the source term in \eqref{eq:r:dimless}. Therefore, we impose
\begin{equation}
	\lim_{x\to\infty}\left(B+\frac{i\pi}{4}\int_b^xH_2^+(x')g(x')x'dx'\right)H_2^-(x)=0,
\end{equation}
so
\begin{equation}
	B=-\frac{i\pi}{4}\int_b^\infty H_2^+(x')g(x')x'dx'.
\end{equation}
This allows to write the full solution as
\begin{multline}
	\bar\Psi(x)=
	\left(
		A - \frac{i\pi}{4}\int_a^xH_2^-(x')g(x')x'dx'
	\right)
	H_2^+(x)	\\
		-\frac{i\pi}{4}H_2^-(x)\int_x^\infty H_2^+(x')g(x')x'dx'
\end{multline}

In order to study the behavior around $x=0$, as well as to make explicit the real and imaginary parts of the solution, we decompose the Hankel functions following \eqref{eq:hankel2JY}. We define $A=\beta+i\gamma$, where $\beta,\gamma\in\mathbb R$, and 
\begin{equation}
	\bar I_Z(x_1,x_2)=\int_{x_1}^{x_2} Z_2(x')g(x')x'dx',
\end{equation}
where $Z$ is either $J$ or $Y$. This allows us to write the solution in its current form as $\bar\Psi(x)=\Re\bar\Psi(x)+i\Im\bar\Psi(x)$, with
\begin{subequations}
	\begin{multline}
		\Re\bar\Psi(x) = \left( \beta  - \frac\pi4\bar I_Y(a,\infty) + \frac\pi2 \bar I_Y(x,\infty) \right)J_2(x)
		                \\- \left(\delta-\frac\pi4\bar I_J(a,\infty)+\frac\pi2\bar I_J(x,\infty)\right)Y_2(x),
	\end{multline}
	\begin{multline}
		\Im\bar\Psi(x) = \left( \beta - \frac\pi4\bar I_Y(a,\infty)  \right)Y_2(x) \\+ \left(\delta-\frac\pi4 \bar I_J(a,\infty)\right)J_2(x).
	\end{multline}
\end{subequations}

The imaginary part has a limit
\begin{equation}
	\lim_{x\to0}\Im\bar\Psi(r)=\left( \beta - \frac\pi4\bar I_Y(a,\infty)\right)\lim_{x\to0}Y_2(x).
\end{equation}
Since $Y_2$ diverges at the origin, we impose
\begin{equation}
	\beta=\frac\pi4\bar I_Y(a,\infty).
\end{equation}
This yields a limit for the real part
\begin{multline}
	\lim_{x\to0}\Re\bar\Psi(r)=\frac\pi2\lim_{x\to0}\bar  I_Y(x,\infty)J_2(x)
   	\\- \left(\delta-\frac\pi4\bar I_J(a,\infty)+\frac\pi2\bar I_J(0,\infty)\right)\lim_{x\to0}Y_2(x).
\end{multline}
For low $x'$, the integrand of $\bar I_Y$ diverges as $g(0)/x$. Similarly, $J_2(x)\sim x^2$. Thus, the first limit vanishes unless $g(x)$ diverges faster than $1/x^2$ around the origin. Avoiding such cases (as we do with all choices in this work), finiteness at the origin requires
\begin{equation}
	\delta-\frac\pi4\bar I_J(a,\infty)=-\frac\pi2\bar I_J(0,\infty).
\end{equation}
We can finally write the real and imaginary parts as
\begin{subequations}
	\begin{align}
		\Re\bar\Psi(x) &= \frac\pi2 \bar I_Y^\infty(x)J_2(x)
		                 + \frac\pi2\bar I_J^0(x)Y_2(x),\\
		\Im\bar\Psi(x) &= -\frac\pi2\bar I_J^{\mathbb R_+}J_2(x).
	\end{align}
\end{subequations}
where $\bar I_Y^\infty(x)=\bar I_Y(x,\infty)$, $\bar I_J^0(x)=\bar I_J(0,x)$ and $\bar I_J^{\mathbb R_+}=\bar I_J(0,\infty)$.

Finally, the physical solution may be written as 
\begin{multline}
	\psi(t,r,\theta)=\Im\tilde\psi(r,t,\theta)=\Im\bar\Psi(kr)\cos(2\theta-2\omega_0t)\\+\Re\bar\Psi(kr)\sin(2\theta-2\omega_0t)\\
	=\frac\pi2\left[ \bar I_Y^\infty(kr)J_2(kr) + \bar I_J^0(kr)Y_2(kr)\right]\sin(2\theta-2\omega_0t)
	\\-\frac\pi2\bar I_J^{\mathbb R_+}J_2(kr)\cos(2\theta-2\omega_0t).
\end{multline}
The rotation and angular velocity fields are easily obtained from this as $\phi(t,r,\theta)=-\Delta\psi(t,r,\theta)$ and $\Omega(t,r,\theta)=\dot\phi(t,r,\theta)$. The results of this are shown in \eqref{eq:phi}-\eqref{eq:omega-2}, where
\begin{equation}
	I_Z(x_1,x_2)=-\frac{\mu k^2}{\iota}\bar I_Z(x_1,x_2)=\int_{x_1}^{x_2}Z_2(x')F(x'/k)x'\mathrm dx'.
\end{equation}

\end{document}